# High-gravity Spreading of Liquid Coatings on Wetting Flexible Substrates

Chen Yang[1], Adam Burrous[1], Jingjin Xie[1], Hassan Shaikh[1], Akofa Elike-Avion[1], Luis Rojas[1], Adithya Ramachandran[1], Wonjae Choi[2], and Aaron D. Mazzeo[1*]

[1]Department of Mechanical and Aerospace Engineering, Rutgers University
[2]Department of Mechanical and Aerospace Engineering, UT Dallas
*corresponding author: aaron.mazzeo@rutgers.edu

**Abstract**

This work describes a mechanical approach for manipulating the capillary length and spreading of liquid coatings on flexible substrates with high gravity. Experimental verification in the literature has focused on cases under standard gravity on earth, and to the authors' knowledge, this work is the first to explore its relevance to spreading puddles under high gravity. By using centrifugation with a high-density liquid base underneath a coated substrate, it is possible to apply acceleration normal to a substrate to decrease the capillary length and increase the rate of spreading. Due to the nature of centrifugation, this method works primarily on flexible substrates, which bend with a curvature that conforms to a contour of uniformly distributed centrifugal acceleration. With high gravity of 600 g applied, the capillary length reduces by a factor of 24.5, and the spreading shifts from "transient spreading" between the surface tension-driven and gravity-driven regimes to the gravity-driven regime. Experimental results show that high gravitational acceleration will enhance the rate of spreading such that a puddle, which would require 12 hours under standard gravity on earth to go from an 8-μl droplet to a 40-μm thick puddle, would require less than 1 minute under 600 g. Overall, this work suggests that previously derived expressions for gravity-driven spreading of puddles under earth's standard gravity extend to predicting the behavior of puddles spreading on smooth, wetting substrates exposed to more than 100 g's of acceleration.

**Introduction**

There are numerous studies for the spreading of liquids on solid surfaces [1]–[6], however, how high gravity influences the spreading of droplets/puddles on surfaces is a topic that has not received significant attention. Lopez et al. [1] developed equations for gravity-driven spreading of a liquid on a solid substrate under standard gravity on earth in 1976. Tanner [5] studied spreading dominated by surface tension in 1979. De Gennes, Oron et al. and Leger et al. also gave thorough and detailed summaries for spreading [2], [6], [7]. Brochard-Wyart et al. [3], Redon et al. [4], and Cazabat and Stuart [8] studied spreading of "heavy" droplets and further illuminated the transition between surface tension-driven and gravity-driven droplets. While there have been efforts to create thin coatings in spinning cylindrical vessels with high gravity [9]–[11], there are no experimental or numerical studies that have focused on the spreading rate of puddles under high gravity.

The physics for complete spreading have dependence on the capillary length – ratio between surface tension and potential energy and the ratio between viscosity and potential energy. This work uses centrifugation to create a high gravity field normal to a flexible substrate. It shortens the capillary length and magnifies the effect of gravity during spreading. The experimental results from this work suggest that the scaling relationships by Lopez et al. [1] with respect to time are still relevant for puddles spreading in the gravity-driven regime.

In this work, we applied high gravity (600 g) using centrifugal processing [12], [13] to deposited quantities of liquids on completely wetting surfaces to make the first reported measurements for spreading puddles under high gravity. In order to expose droplets/puddles to high gravity, we provided centrifugal acceleration normal to a flexible substrate. The thin flexible substrates, which sat on a dense liquid, bent with a radius that was approximately equal to the distance from the axis of rotation. In this way, the acceleration normal to the substrate was uniform. After centrifugation, an imaging system took a snapshot of the puddle on the substrate. We then estimated the thickness for this nonvolatile material by dividing its volume by the measured area.

In the case of complete wetting, a central drop and a precursor film together form a drop/puddle. The focus in this paper is on the central drop as it includes the coated area of interest with thickness greater than 1 μm [2]. In general, there are three regimes of spreading due to different driving forces for complete wetting as shown in Figure 1. In each of these regimes, the radius of a puddle increases as a function of time. For the surface tension-driven regime, the radius grows with $t^{1/10}$ [2], [5], [6] as shown in

$$R(t) = V^{3/10} \left( \frac{\gamma_{LV} t}{\eta} \right)^{1/10}, \tag{1}$$

where the symbols represent the spreading time ($t$), the horizontal radius ($R$), the liquid-vapor surface tension ($\gamma_{LV}$), the volume ($V$), and the kinematic viscosity ($\eta$) of the spreading droplet. For sufficiently large drops in the gravity-driven regime, the radius grows with $t^{1/8}$ [1], [2]. In the final stage of spreading dominated by "long-range" forces, there is not a simple closed-form relationship.

For gravity-driven spreading, Lopez et al. derived a governing equation for an axisymmetric drop in the case of complete wetting [1], and the equation can be reorganized into three different forms depending on the experimental quantity of interest:

$$R(t) = \left[ 0.136 \frac{V^3 \rho g}{\mu} t \right]^{\frac{1}{8}} \tag{2}$$

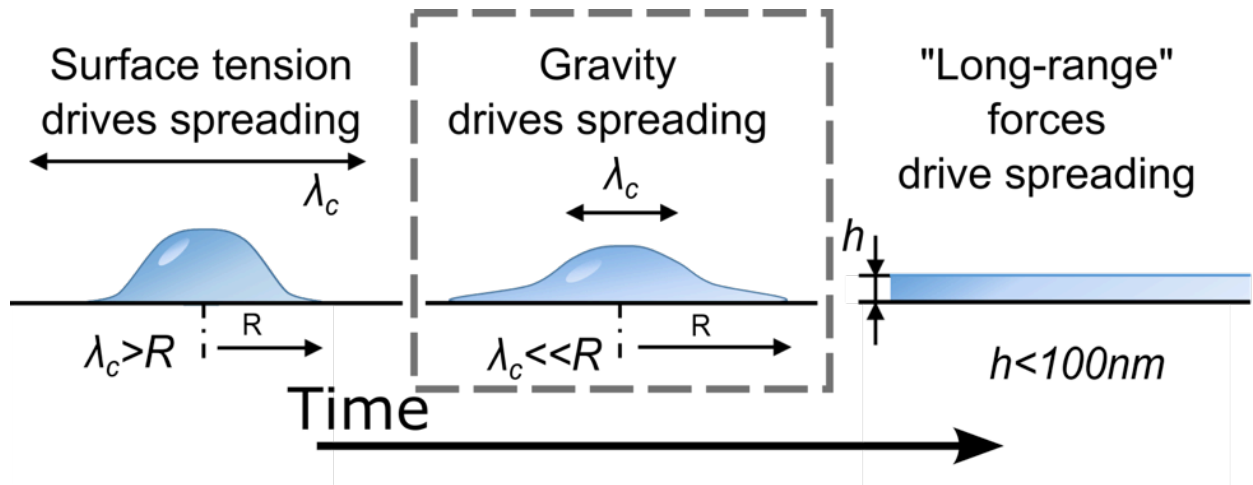

**Figure 1: Three general regimes for spreading of completely wetting puddles: surface tension-driven spreading (radius of puddle is less than capillary length), gravity-driven spreading (radius is much greater than capillary length), and spreading driven by "long-range forces" (thickness is less than 100 nm). This paper focuses on the gravity-driven regime with the use of centrifugation.**

$$A(t) = \pi (R(t))^2 = \pi \left[ 0.136 \frac{V^3 \rho g}{\mu} t \right]^{\frac{1}{4}} \qquad (3)$$

$$h(t) = \frac{1}{\pi} \left( \frac{\mu V}{0.136 \rho g} \right)^{1/4} t^{-1/4} \qquad (4)$$

In Equations 2-4, the symbols represent the spreading time ($t$), the horizontal radius ($R$), the area ($A$), the applied gravity (g), the thickness ($h$), the volume ($V$), and the dynamic viscosity ($\mu$) of the spreading droplet.

As we use centrifugation to increase the rate of spreading/coating, we have an interest in understanding the bounds of the gravity-driven regime, which we will describe in more detail. The transition between surface tension-driven and gravity-driven regimes is dependent on the capillary length ($\lambda_c$) – a ratio between the liquid-vapor surface tension ($\gamma_{LV}$) and potential energy (the product of density ($\rho$) and gravity ($g$)):

$$\lambda_c = \sqrt{\frac{\gamma_{LV}}{\rho g}} \qquad (5)$$

Under normal gravity, a liquid-vapor surface tension of silicone oil is around 20 mN/m, which results in a capillary length of 1.5 mm. A hemispherical droplet with a radius of this capillary length would have a volume of 7.1 μl. Under 600 g, which is the typical acceleration applied in this work, the capillary length decreases to 61 μm and has a corresponding volume of 0.5 nl.

Brochard-Wyart et al. [3], [4] showed that in the transition between the gravity-driven and surface tension-driven regimes, the spreading does not abruptly switch to the gravity-driven regime for large droplets studied by Lopez et al. [1]. The spreading in this intermediate regime does not obey a relationship with a fixed exponent (i.e., 1/10 for the surface tension-driven regime and 1/8 for the gravity-driven regime). There is a critical size of the droplet $R_c$ defined as follows, which marks the beginning of gravity-driven spreading as described by Lopez et al.

$$R_c \cong 4L\lambda_c \qquad (6)$$

$L$ in the equation is a logarithmic factor [6]. To date, $L$ is only determinable through observation of drop profiles and fitting. Redon et al. [4] found an $L$ of ~3 for a system of polydimethylsiloxane (PDMS) and silica surface, which gave an approximate critical radius of $12\lambda_c$. Cazabat and Stuart [8] also demonstrated that the critical radius was dependent on the volume of the deposited droplet.

Application of high gravity reduces the capillary length (see Equation 5). Reducing the capillary length will enlarge the effect of gravity on small drops (see Figure 1) by shifting the transition between surface tension-driven spreading and gravity-driven spreading as described by $\lambda_c$ and $R_c$. With this shifted transition, gravity plays a dominant role on smaller puddles of a given volume and accelerates the spreading process. Similarly, introduction of small droplets to a substrate under high gravity means that gravity facilitates the flattening of micro puddles, which would otherwise take hours or days to thin under surface tension-driven spreading. In addition to spreading being proportional to $t^{1/8}$ instead of $t^{1/10}$, the leading coefficient in Equations 2-4 can change by a factor greater than 2 for 600 g of applied acceleration, while the leading coefficient in surface tension-driven spreading given in Equation 1 remains unchanged.

Due to the strong nonlinearity of Lopez's relationships (Equation 2-4), it may not appear practical to coat large surfaces by manipulating gravity. For example, doubling the area or halving the thickness for a drop with a fixed volume of liquid requires the spreading time to be 16 times longer. Nonetheless, by increasing gravity with centrifugation and by understanding that the proportionality between time and coated "area" is linear, high-gravity spreading of puddles merits further exploration as a potential manufacturing process. As shown in Equation 9, assume $V_2 = nV_1$ and $t_2 = nt_1$, there is a linear scaling between area and time.

$$A_1(V_1, t_1) = \pi \left[ 0.136 \frac{V_1^3 \rho g}{\mu} t_1 \right]^{\frac{1}{4}} \quad (7)$$

$$A_2(V_2, t_2) = \pi \left[ 0.136 \frac{V_2^3 \rho g}{\mu} t_2 \right]^{\frac{1}{4}} \quad (8)$$

$$\frac{A_2(V_2, t_2)}{A_1(V_1, t_1)} = \frac{\left[ (nV_1)^3 (nt_1) \right]^{1/4}}{\left[ (V_1)^3 (t_1) \right]^{1/4}} = n \quad (9)$$

With centrifugal acceleration applied normal to substrates, material processes, with further development, may offer precise control over the area, thickness, and uniformity of coatings. For processes or devices requiring dielectric and conductive coatings, which often lose expensive precursors in batch-based spin coating, this method may reduce their waste significantly. We anticipate that the physical understanding of how puddles spread under high gravity will have relevance to the manufacture of single-layer coatings for protecting surfaces (e.g., digital displays, sensors, and keypads) and multilayer coatings or stacks of multiple materials (e.g., solar cells, flexible electronics, and dielectric elastomers).

**Experimental Design**

*Spreading under Standard Gravity*

This work focuses on high-gravity spreading (see SI Figure 1) and how spreading differs from that under standard gravity on earth. In order to apply earth's gravity uniformly to puddles, we first created a level and flat base by pouring Mold Star 30 (Smooth-On, Inc.) and curing it in a Petri dish fixed to a table. On top of the cured Mold Star, we laid a 50 μm-thick sheet of polyethylene terephthalate (PET) - length and width of 37 mm - with a manufacturer-specified surface tension of 45 mN/m (Mylar WC, Dupont Teijin Films). We also set an SLR camera (Nikon D7100) above the Petri dish to take photos of the spreading liquid in the Petri dish (see SI Figure 2). After making sure the surface of the PET was clean and there were no air bubbles trapped between the PET and the Mold Star, we dispensed one droplet of 1000 cSt silicone oil (Sigma-Aldrich Corporation) on the surface of the PET using a high-precision dispenser (Performus II, Nordson EFD). To explore the regime of gravity-driven spreading, we ran experiments with volumes of 8 μl and 80 μl. In our experiments, we took images at 11 timestamps, which were evenly distributed on a logarithmic scale from 1 minute to 10 minutes.

*Spreading under high gravity*

To achieve uniformly distributed acceleration normal to the substrate on which a puddle spread, we first floated a flexible substrate (i.e., the same type of polyethylene terephthalate (PET) used previously under standard gravity with a specific density of 1.4 g/cm$^3$) on a high-density fluid (sodium polytungstate from GeoLiquids, Inc. with a specific density of 3.1 g/cm$^3$) in a trough/Petri dish as shown in Figure 3 a. Then, the flexible substrate received a liquid droplet, and the Petri dish, high-density liquid, flexible substrate, and dispensed droplet, went into a swinging bucket of a bench-top centrifuge (CR4-12 Refrigerated Centrifuge, Jouan/Thermo as shown in SI Figure 3). As the centrifuge spun up to a set speed (2000 rpm, ~ 600 g), the bucket tilted and went nearly horizontal as shown in Figure 3 b and c. The surface of the high-density fluid conformed to a cylindrical radius of curvature approximately 133 mm from the axis of rotation to reach a steady state of uniformly applied centrifugal acceleration. During spin up, the high-density fluid sitting in the trough sloshed very little with an appropriate balance of Euler and normal accelerations. While the centrifuge was spinning at a set speed, uniform high gravity normal to the surface of the flexible substrate acted on the spreading puddle. In our experiments, we ran one sample at each timestamp because we had to stop the centrifuge to take an image of the result with a camera. The camera (Basler acA2040-180km) took images with a collimated LED light (BL0404, Advanced Illumination) transmitting light through the backside of the sample (see SI Figure 4).

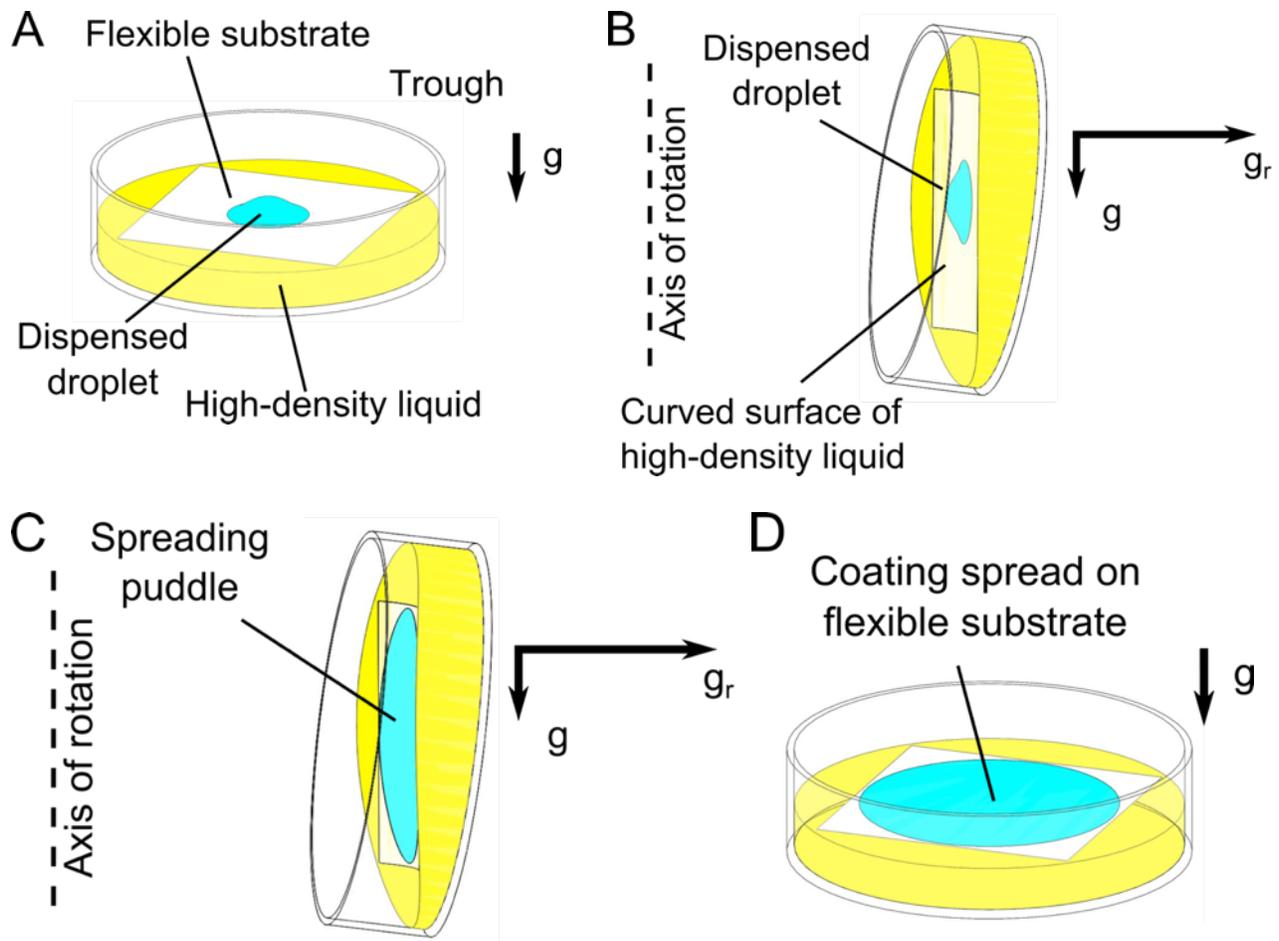

**Figure 3: Basic overview of the experimental setup for high-gravity spreading/coating. (a) Trough of high-density liquid with a floating flexible substrate on its surface. (b) Spinning trough/Petri dish under high gravity with a droplet spreading on the flexible substrate. The slight tilt of the trough is from the swinging bucket not quite going horizontal. (c) Same as b after elapsed time. (d) Resulting liquid coating spread on the flexible substrate.**

*Image processing*

We used the MATLAB Image Processing Toolbox to post process the images taken in the experiments. For convenience, we wrote a MATLAB script to measure automatically the areas of the puddles and estimate average thicknesses of the puddles on the substrates (see Supplementary Information). In our script, we first cropped images to a desired size. Then, we enhanced the contrast of the image to highlight the puddle. Next, selecting pixels brighter than a threshold and connecting the selected pixels to neighboring pixels resulted in a few highlighted areas. In the final step, we chose the largest area and used the built-in convex-hull algorithm to fill the area of the highlighted circle or ellipse. We used this set of processes to find the area of the top surface of the puddles. By considering the shapes of the puddles to be thin circular plates (volume = area x height), we estimated the thicknesses.

**Results and Discussion**

In studying the high-gravity spreading of silicone oil on PET, we used centrifugation as previously described to obtain images of spreading puddles as shown in Figures 4 and 5. We then matched our data under standard gravity to Lopez's equations and made further comparisons to data take from high-gravity experiments. In this work, we only compared our measured data to Lopez's predictions in the gravity-driven regime because we increased gravity and selected sufficiently large volumes of droplets to ensure their effective hemispherical radius would be larger than the critical radius. To align our data and Lopez's prediction, we used the first measured data point associated with 60 seconds of spreading or spinning as a reference, as the initial condition in the experiments did not exactly match that assumed as the initial geometry in Lopez's work [1]. For instance, there is a short period of time during extrusion of the droplet from the syringe to the surface when the radius of a droplet is less than the capillary length. In addition, there is a finite length of time required to dispense droplets (e.g., 0.35 s for 8-μl droplets and 3.5 s for 80-μl droplets), in which the droplets are spreading and growing at the same time. Also, Lopez's solution only applies after the drop spreads to an area much larger than its initial one to ensure the validity of similarity solution. For high-gravity spreading, there are additional factors that may cause this mismatch, which we will discuss later in detail.

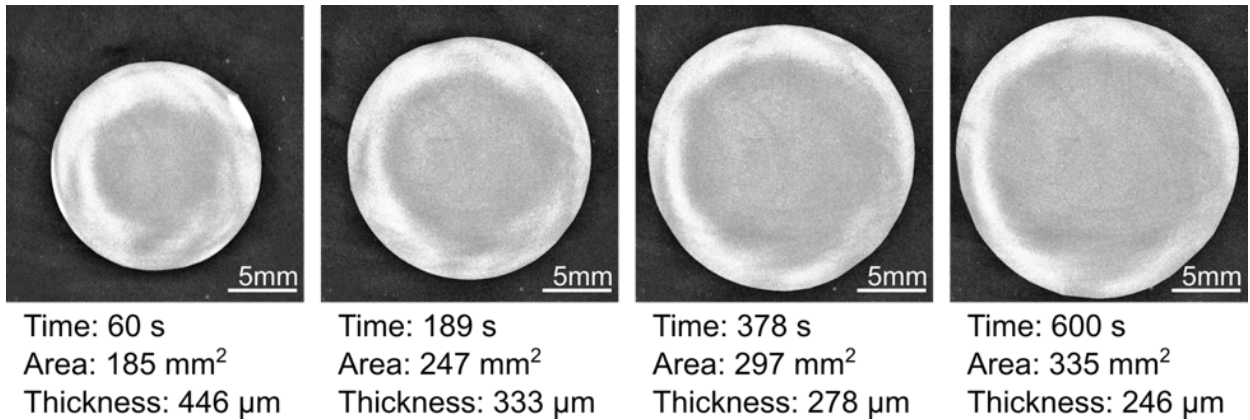

| Time: 60 s | Time: 189 s | Time: 378 s | Time: 600 s |
| Area: 185 mm² | Area: 247 mm² | Area: 297 mm² | Area: 335 mm² |
| Thickness: 446 μm | Thickness: 333 μm | Thickness: 278 μm | Thickness: 246 μm |

**Figure 4: A puddle of silicone spreading on a PET under 1 g. The dispensed volume was 80 μl, the density was 960 kg/m$^3$, and the viscosity was 1 Pa•s. Images processed using customized algorithms in the MATLAB Image Processing Toolbox show the puddle at different time of spreading. Puddles with this volume spread in the gravity-driven regime.**

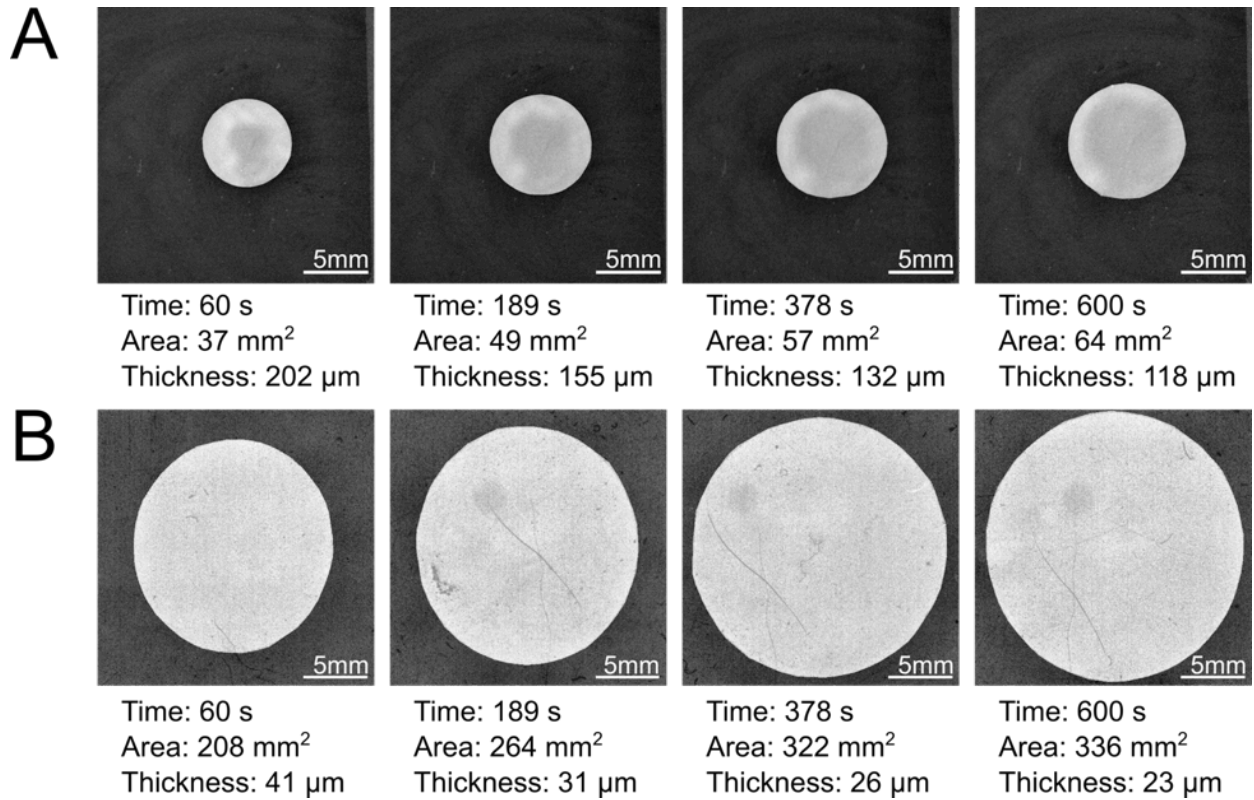

Figure 5: (a) A puddle of silicone spreading on a sheet of PET under 1 g. (b) Different puddles spun for the specified periods of time at 2000 rpm with 600 g normal to the PET. In both cases, the dispensed volume was 8 µl, the density was 960 kg/m3, and the viscosity was 1 Pa•s. Images processed using customized algorithms in the MATLAB Image Processing Toolbox show the puddles at different times of spreading. Puddles with this volume spread in an intermediate regime under 1 g and spread in the gravity-driven regime under 600 g.

To align the data and Lopez's model, the equations for calibration are:

$$\log_{10}(h) = m \times \left(\log_{10}(t) - \log_{10}(60) + \log_{10}(t_i)\right) - m \times \log_{10}\left(\frac{\mu V}{0.136 \rho g \pi^4}\right)$$

$$= m \times \log_{10}(t) + C, \tag{8}$$

where

$$C = m\left[-\log_{10}(60) + \log_{10}(t_i) - \log_{10}\left(\frac{\mu V}{0.136 \rho g \pi^4}\right)\right] = m \log_{10}\left(\frac{0.136 \rho g \pi^4 t_i}{60 \mu V}\right). \tag{9}$$

Table 1 shows calibrating parameters and fitted equations in three different sets of experiments. The slope $m$ in Lopez's equation is -0.25 from Equation 4, $C$ is the calibrated/fitted parameter after lining up Lopez's prediction with our experimental results, $t_i$ is the corrected initial time offset from Lopez's prediction to the initial data point taken after 60 s of spinning/spreading, slope $m'$ and $C'$ under "Fitted Equation" are the fitted parameters from experimental results. Figures 6 and 7 present plots for the data, fittings and predictions.

In Redon et al.'s work [4], they observed a critical radius to delineate the start of the

**Table 1. Calculated parameters after aligning the experimental data with Lopez's equation and fitted parameters from results. The only adjusted parameters for the least squares regression were *m* and *C*.**

| Experiment | Lopez's Equation | | | | Fitted Equation | |
|---|---|---|---|---|---|---|
| | *m* | *C* (log$_{10}$(μm)) | $\frac{\mu V}{0.136 \rho g \pi^4}$ (μm$^4$/sec) | $t_i$ (sec) | *m'* | *C'* (log$_{10}$(μm)) |
| 1 g (~80 μl) | -0.25 | 3.08 | 6.16 x 10$^{11}$ | 17.4 | -0.251 | 3.08 |
| 1 g (~8 μl) | -0.25 | 2.76 | 5.99 x 10$^{10}$ | 34.3 | -0.219 | 2.7 |
| 600 g (~8 μl) | -0.25 | 2.05 | 1.03 x 10$^8$ | 37.8 | -0.242 | 2.04 |

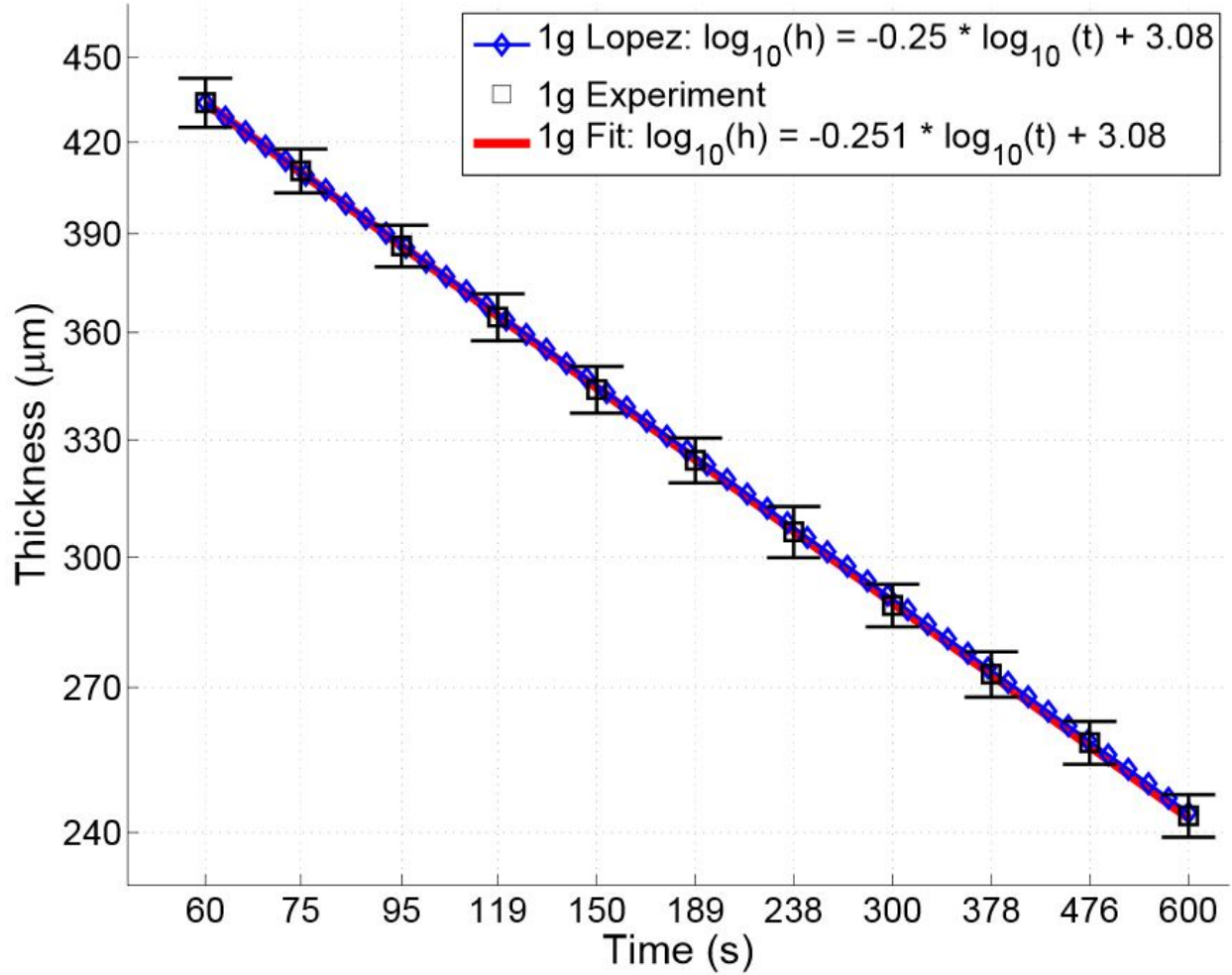

**Figure 6: Theoretical, experimental, and fitted thicknesses with error bars of ±1 standard deviation on either side of the mean for spreading puddles of silicone oil (~80 μl) on a wetting substrate at 1 g over a period of 10 minutes. There were 7 repetitive measurements at each of 11 experimental timestamps used to calculate the mean thickness of spreading puddles.**

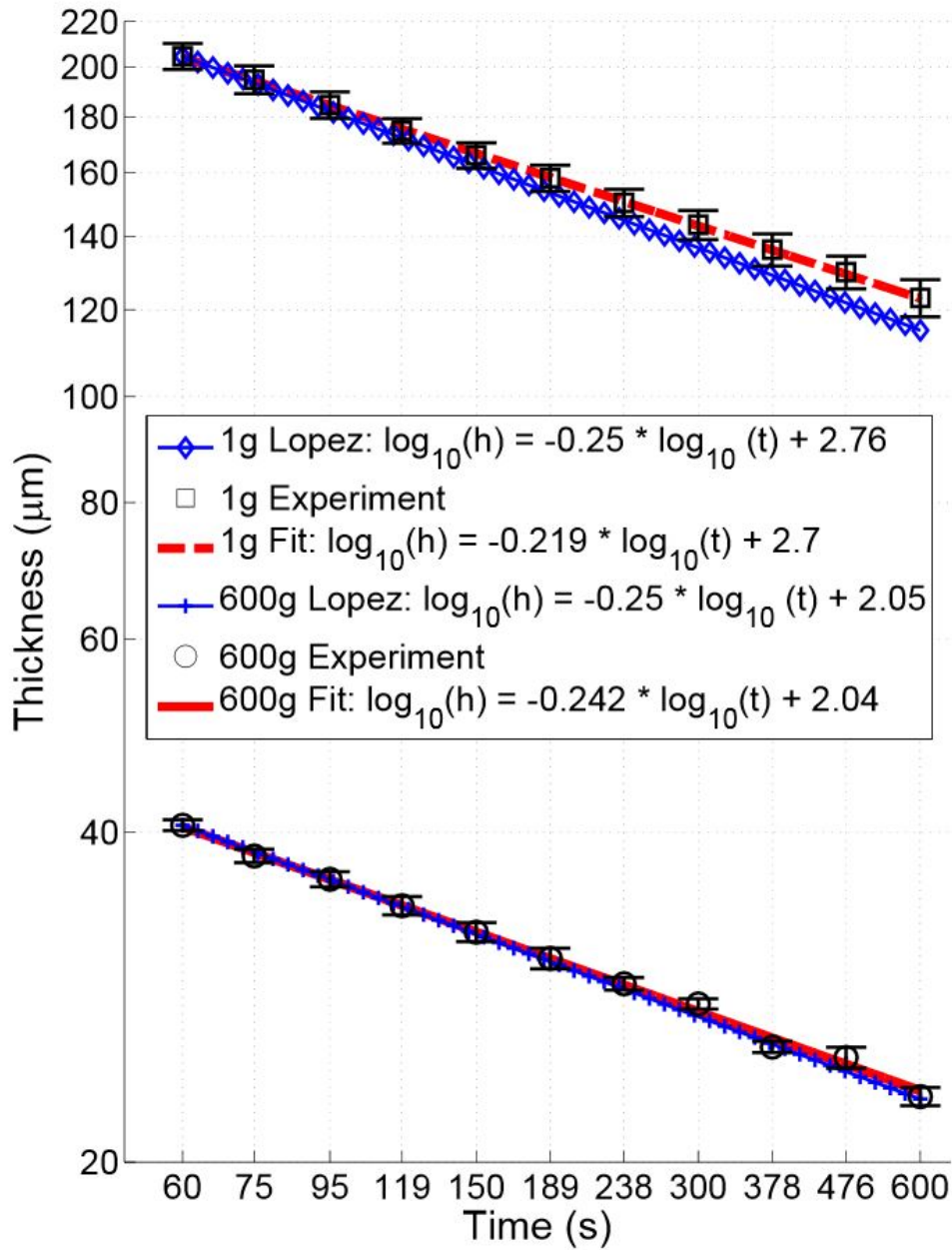

**Figure 7: Theoretical, experimental, and fitted thicknesses with error bars of ±1 standard deviations on either side of the mean values for spreading puddles of silicone oil (~8 μl) on a wetting substrate at 1 g and at high gravity of 600 g over a time frame of 10 minutes. There were 10 repetitive experiments at each timestamp for 1 g and 7 repetitive experiments at each timestamp for 600 g.**

gravity- driven regime, as described by Lopez et al., $R_c \approx 12\lambda_c$ for PDMS on silica. To measure spreading beyond this intermediate regime, we used a large puddle of silicone oil (~80 µl) as shown in Figure 4 to verify Lopez's relation under standard gravity. There were seven puddles with measurements at each timestamp, which spread from start to 10 min (unlike the experiments under high gravity that stopped at each timestamp to take an image). Modeling the puddle as a thin circular plate, we estimated the radius from the area of the highlighted region using previously described scripts in MATLAB. The radius was ~7.7 mm ($\sim 5\lambda_c$) at 1 min after dispensing and reached ~10.3 mm ($\sim 7\lambda_c$) at 10 min. Figure 5 shows the thickness calculated by preserving the volume and dividing it by area measured. The experimental results agree with Lopez's equations, which is that the thickness decreased proportional to $t^{-1/4}$ or the radius increase proportional to $t^{1/8}$. From this result and in contrast to PDMS spreading on silica as described by Redon, et al., we found the critical radius was likely no more than $5\lambda_c$, the radius at the first timestamp of our experiments, for silicone oil spreading on PET.

    Lopez et al. [1] used data from another study [14] to verify their model of spreading in the gravity-driven regime but did not mention the critical radius that separates transient spreading and the gravity-driven regime. For our experiments with silicone oil on PET, we have also observed transient spreadingfor 8-µl drops under standard gravity on earth, which has a rate of spreading ($m = -0.22$) different from the surface tension-driven and gravity-driven regimes. Experiments under standard gravity of 9.8 m/s² with a small droplet of silicone oil (~8 µl) measured the rate of spreading when the spreading was not yet in the gravity-driven regime. There were 10 repetitive measurements on distinct puddles at each timestamp performed using the same protocol to collect the data shown in Figure 4. Figure 5 a shows selected images from

these experiments. Figure 7 shows that the measured results of these experiments with 8-μl drops at 1 g do not agree with Lopez's equations under 1 g as well as the experiments performed with 80-μl drops. Approximating the puddles as cylinders, the radius of the drop at 1 min was ~3.5 mm ($\sim 2.3\lambda_c$) and ~4.5 mm ($\sim 3\lambda_c$) at 10 min. The fitted curve for this experiment shows an exponent of -0.219, which is an intermediate value (i.e., -0.2 would correspond to the surface tension-driven regime and -0.25 would correspond to the gravity-driven regime) for the power scaling of thickness over time. Given the difference between the fitting and Lopez's equation, these results suggest that spreading occurred in the intermediate regime and that the critical radius was greater than $3\lambda_c$.

Using the protocol described for high-gravity spreading, Figure 5b shows representative results. There are 7 measurements at each of 11 timestamps from a total of 77 puddles spread in our centrifuge. Comparing the spreading of a small droplet of silicone oil (~8 μl) into a puddle under 1 g to puddles spun at 600 g for periods of time ranging up to 10 minutes in Figure 7, it is apparent that a puddle exposed to high gravity spreads more quickly into a larger, thinner puddle than one spreading under standard gravity on earth.

In Figure 5b, the images have a slightly elliptical shape as the distance from the top to the bottom is longer than that from left to right. One proposed reason is that when the centrifuge began to spin, the centrifugal acceleration had an outwardly projected component tangent to the substrate before the swinging bucket reached its nearly horizontal position. This projection caused the drop to spread a slightly upward and away from the central axis of rotation. We tested this idea by comparing the eccentricities of measured puddles centrifuged for different times. We also ran samples for a short time (~30 sec) and found similar phenomenon with those samples. With the experiments showing no correlation between the time of spinning and eccentricity, the slight eccentricity of the puddles does not appear to result from Euler forces but non-uniform acceleration normal to the substrate in the outward direction as the centrifuge spins up or accelerates to its steady-state speed.

Unlike the results for the small droplet at 1 g, Figure 7 shows the experimental results at 600 g agreeing reasonably well with predictions made by Lopez's equation for gravity-driven spreading (Equation 6). High gravity brought down the capillary length from 1.5 mm at 1 g to 61 microns at 600 g and ensured the spreading of these small droplets passes into the gravity-driven regime quickly above the critical radius (i.e., radius was ~8 mm ($\sim 131\lambda_c$) at 1 min). However, there is still a small deviation between the theoretical and experimental results. One possible reason is that the centrifuge requires 10 s of seconds for acceleration/deceleration to/from a desired speed of rotation (e.g., 20 s to accelerate to 2000 rpm associated with 600 g and 30 s to decelerate back to rest). The reported run times in this work do not include the time for deceleration but do include the time to accelerate to the desired spin speed/centrifugal acceleration. Another reason is that the reported run times also do not count the time of dispensing and transporting the sample to the centrifuge, which usually takes 15 seconds at 1 g. In other words, these fixed additional times required for processing experiments - or variations in such processing times - might have had a more significant impact on enhancing the effective rates for spreading at low times for spinning, which might have lead to a less negative exponent than expected.

By comparing the thicknesses of samples at different gravities, it is significant that puddles at 1 minute under 600 g have much smaller thicknesses than those having spread for 1 min under 1 g. If we extend Lopez's curve of 1 g to a thickness of 40 μm, which is similar to the thickness after 1 minute of spinning under 600 g, the time required for the spreading under 1 g would be approximately 12 hours. Figure 8 shows snapshots of the same puddle spreading under different gravities to provide a direct comparison between high-gravity spreading and standard-gravity spreading. We conducted the experiment of spreading under standard gravity on the imaging platform because we needed to image at 0 min. After five minutes of spreading at 1 g, we then applied high gravity of 600 g, for 5 minutes to enlarge the puddle significantly. Finally, it is apparent that another 30 minutes of spreading under 1 g only changed the area by less than 0.1%.

**Conclusions and Future Directions**

In this work, we analyzed and experimentally tested the concept of high-gravity spreading of droplets/puddles on flexible substrates in a centrifuge through a unique setup with a flexible substrate floating on high-density liquid. By reducing the capillary length under high gravity, we shifted the spreading of small droplets into the gravity-driven regime. By matching experimental results with theoretical equations governing gravity-driven spreading, we found agreement between theoretical and experimental rates of spreading under standard and high gravity. Coating areas under high gravity showed significantly faster rates of spreading than those studied under normal gravity. The introduction of small droplets to a substrate under high gravity similarly means that high gravity facilitates the flattening of micro puddles, which would otherwise take hours or days to thin under surface tension-driven spreading.

This work also outlines the basic principles for a potential manufacturing method to coat flexible substrates with minimal waste. It also outlines a unique platform for the experimental study of spreading puddles at high gravity. As the time for spreading droplets scales with the size of the coated area, high-gravity spreading of single puddles may be applicable in certain scenarios. Future work will explore the interactions and coating of multiple, small droplets for the potentially rapid, large-area coating of flexible substrates. For functional coatings, optimization will also be necessary to freeze or cure high-gravity coatings once they reach an appropriate size. Although not addressed in this work, high-gravity coating also has the potential to affect the spreading of droplets on partially wetting substrates, which are incompatible with conventional coating techniques.

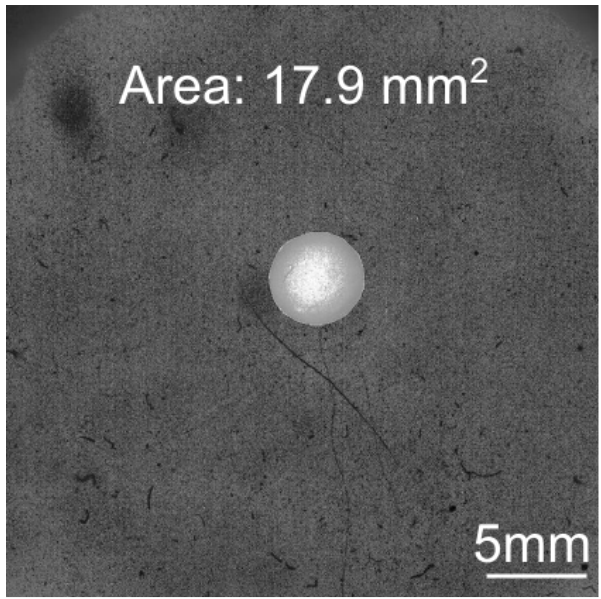
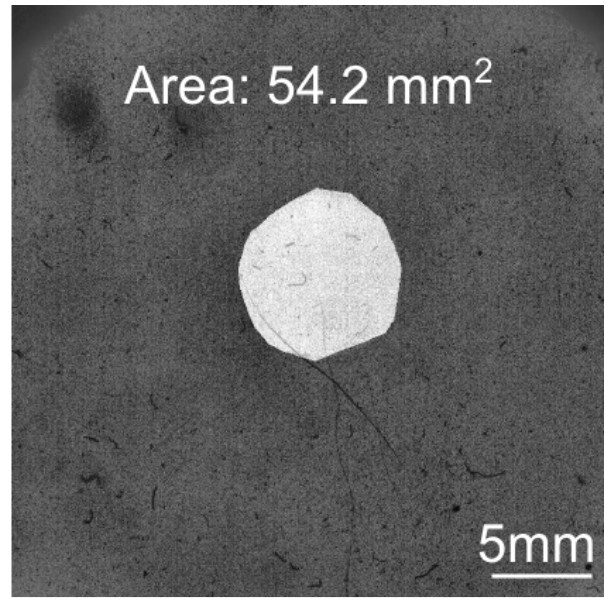
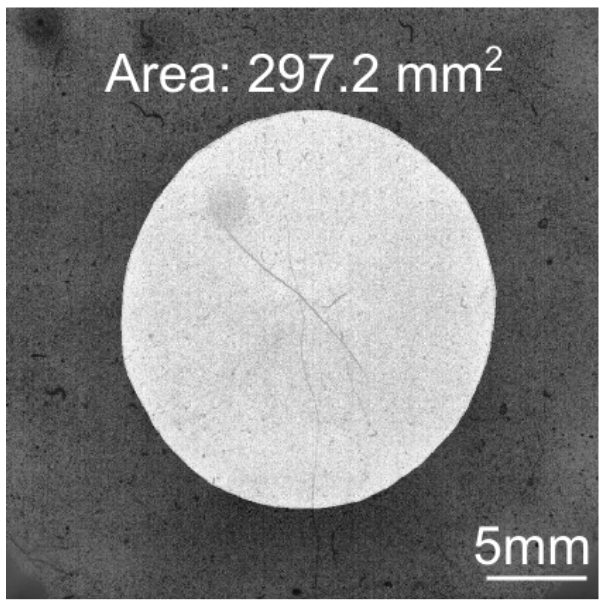
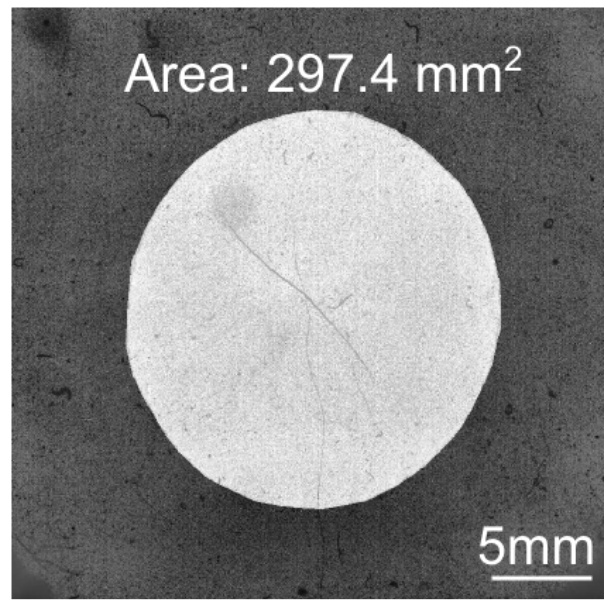

**Figure 8: One drop spreading under different gravities. Spreading for 5 minutes under 600 g significantly enlarged the puddle and spreading for 30 minutes under 1 g after that only changed the area by a small amount (0.2 mm$^2$).**

# High-gravity Spreading of Liquid Coatings on Wetting Flexible Substrates

Chen Yang[1], Adam Burrous[1], Jingjin Xie[1], Hassan Shaikh[1], Akofa Elike-Avion[1], Luis Rojas[1], Adithya Ramachandran[1], Wonjae Choi[2], and Aaron D.Mazzeo[1*]

[1]Department of Mechanical and Aerospace Engineering, Rutgers University
[2]Department of Mechanical and Aerospace Engineering, UT Dallas
*corresponding author: aaron.mazzeo@rutgers.edu

## Supplementary Information

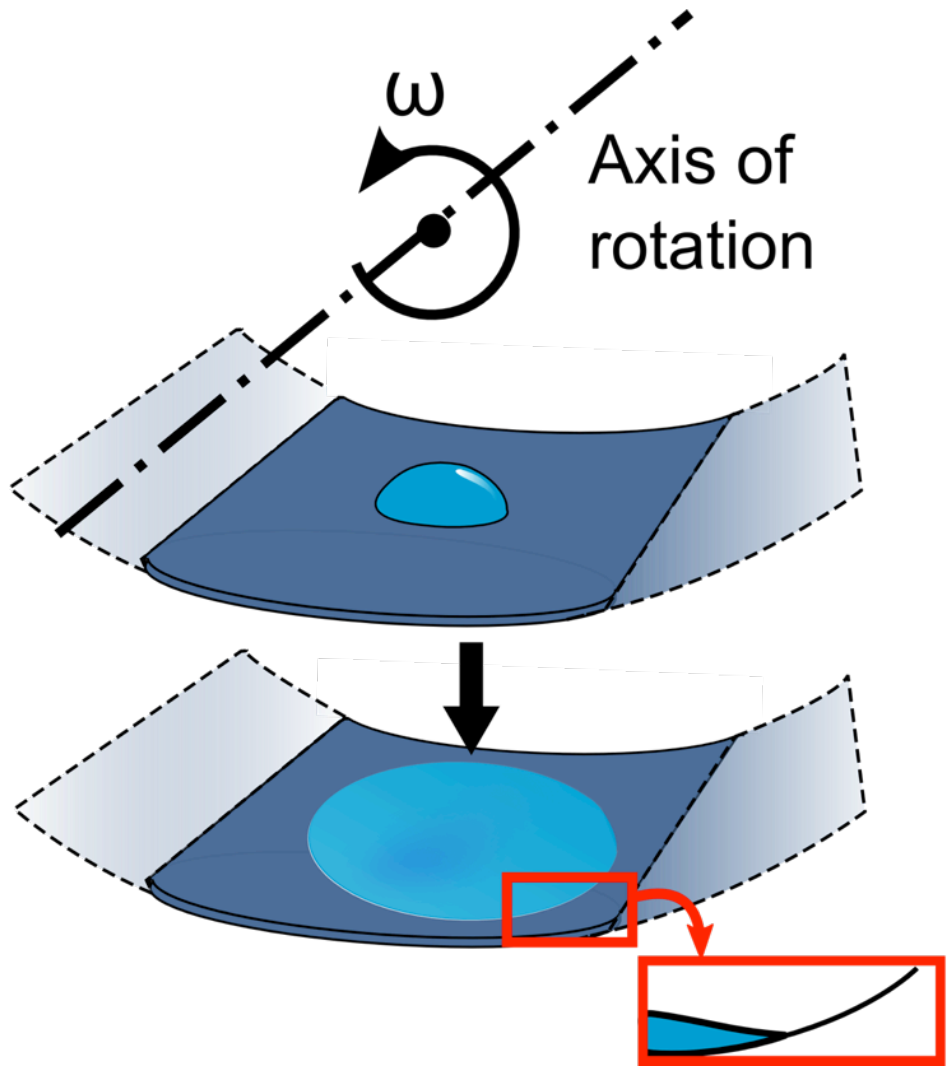

**SI Figure 1: High-gravity coating of a completely wetting puddle on a flexible substrate，which spreads radially outward with the edge having a small advancing contact angle. The flexible substrate sits on a dense liquid (not shown) and bends to conform to the curvature determined by the radial distance from the axis of rotation.**

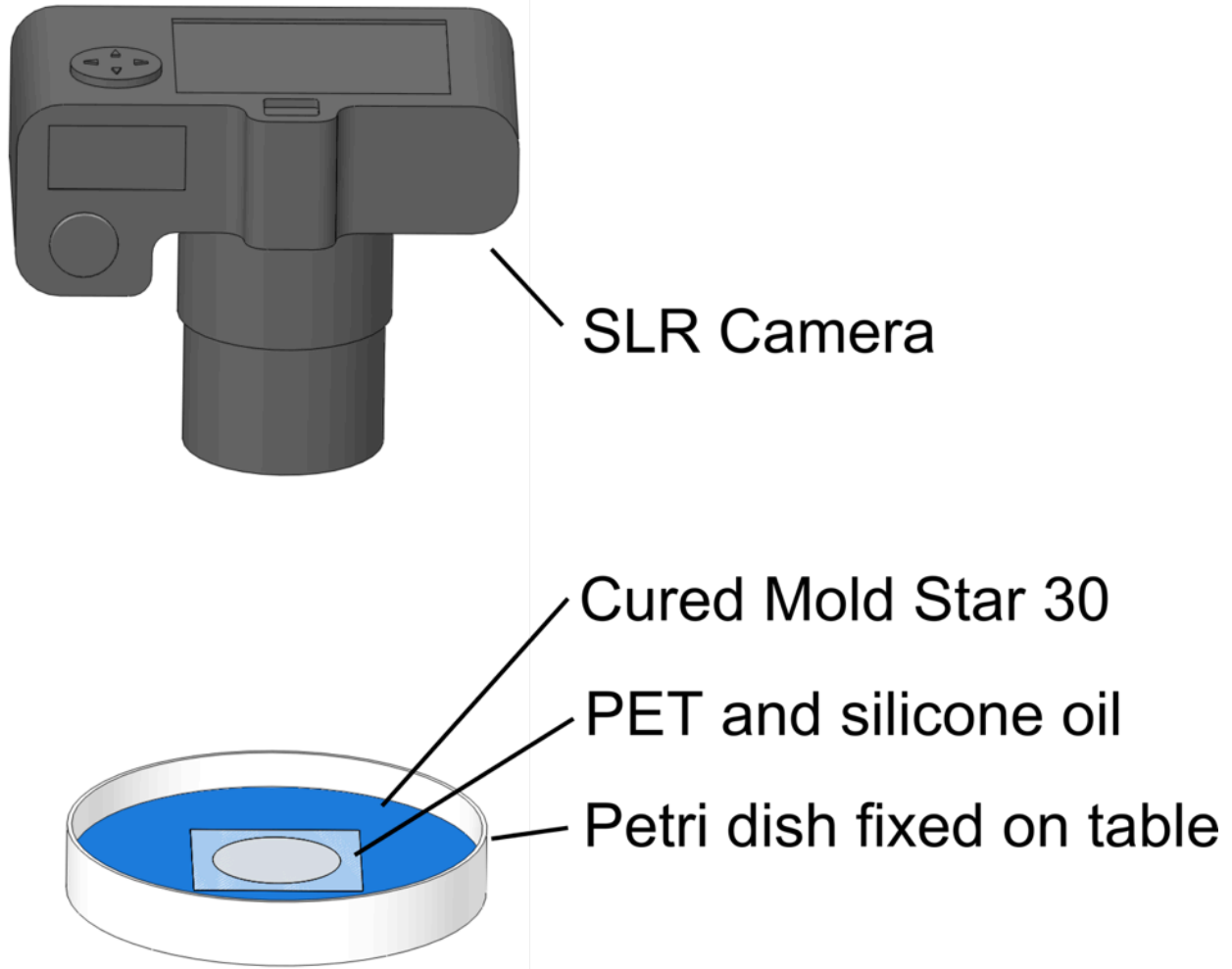

**SI Figure 2: Experimental platform of liquid spreading under standard gravity on earth. The liquid spreads uniformly on the level and flat PET. An SLR takes 11 images at timestamps logarithmically distributed over a period of from 1 min to 10 min to measure the size of the spreading puddle of silicone oil.**

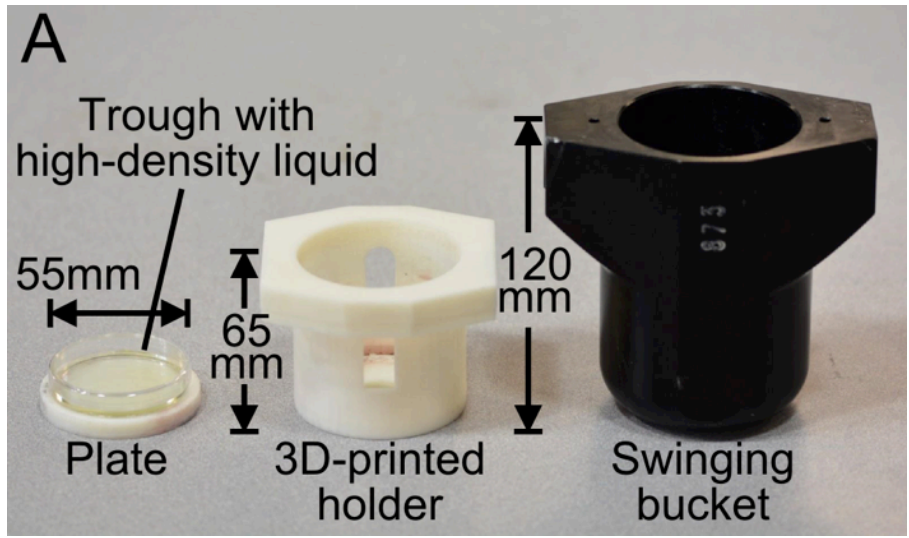
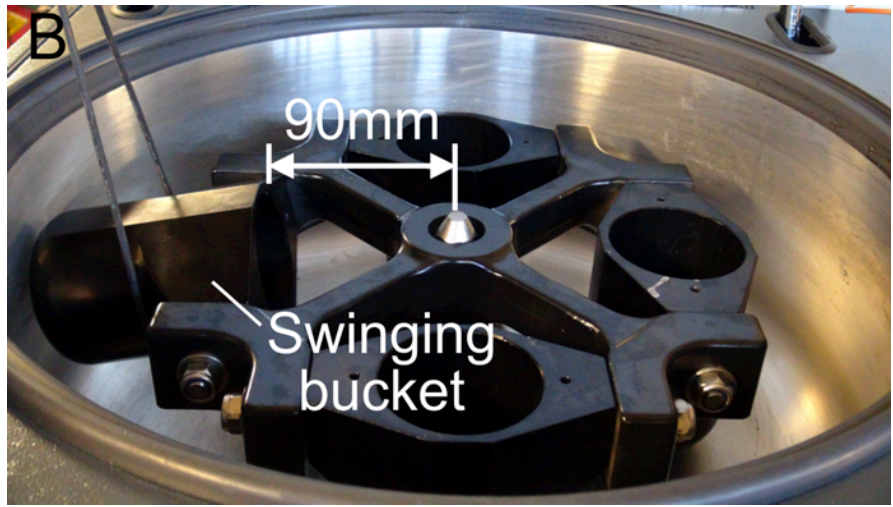

**SI Figure 3: The experimental setup for high-gravity coating.** (a) The trough (Petri dish) holds the high-density fluid, substrate, and coating droplet. The plate goes in the holder of the swinging bucket, which spins in the centrifuge. (b) While spinning within the centrifuge, the bucket swings up to a nearly horizontal position.

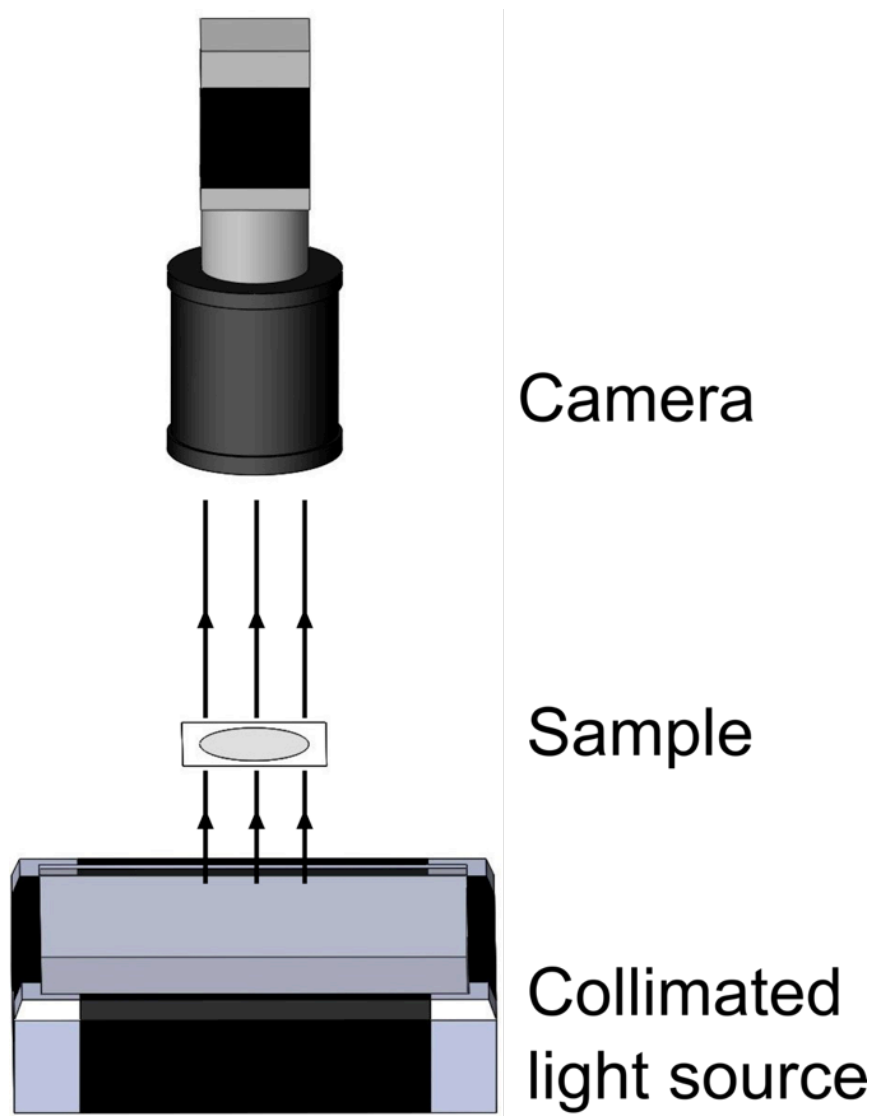

**SI Figure 4: Schematic drawing of imaging platform for centrifuged samples. Measurements on samples occurred after spinning for set periods of time. The source transmits collimated light through the underside of the substrate and puddle to highlight the image received by the camera.**

## MATLAB code for image processing (brief description included in experimental design)

```matlab
clear all
close all

Conversion = 1031/0.037;

a1=imread('1.png');

figure,imshow(a1);
imwrite(a1,'1.jpg');

b=size(a1);
cenrow=b(1,1)/2;
cencol=b(1,2)/2;
crop=imcrop(a1,[cencol-500,cenrow-500,1000,1000]);

figure,imshow(crop);
imwrite(crop,'2.jpg');

%% Contrast Enhancement
crop = adapthisteq(crop);
crop_col = crop(:);
crop_mean = mean(crop_col);
crop_col_double = double(crop_col);
crop_std = std(crop_col_double);
crop_min = crop_mean - crop_std;
crop_max = crop_mean + crop_std;
for i = 1:1000
    for j = 1:1000
        if crop(i,j)-crop_min < 0;
            crop(i,j) = 0;
        else if crop(i,j) - crop_min > 0 && crop(i,j) - crop_max < 0;
                crop(i,j) = round((crop(i,j)-crop_min)/(crop_max-crop_min)*256);
            else if crop(i,j) - crop_max > 0;
                    crop(i,j) = 255;
                end
            end
        end
    end
end
crop = adapthisteq(crop);
figure,imshow(crop);
imwrite(crop,'3.jpg');

figure,imshow(crop);
```

```matlab
imwrite(crop,'4.jpg');

temp=zeros(1000,1000,3);
%% Pixel Selection
for n=1:300
    for m=1:1000
        if crop(n,m)>185;
            temp(n,m,1:3)=255;
        end
    end
end
for n=301:760
    for m=1:725
        if crop(n,m)>190;
            temp(n,m,1:3)=255;
        end
    end
end
for n=301:760
    for m=726:1000
        if crop(n,m)>170;
            temp(n,m,1:3)=255;
        end
    end
end
for n=761:1000
    for m=1:1000
        if crop(n,m)>200;
            temp(n,m,1:3)=255;
        end
    end
end

figure,imshow(temp);
imwrite(temp,'5.jpg');
temp = rgb2gray(temp);

%% Region Properties
cropbw=im2bw(temp);
cropbw1=imfill(cropbw,'holes');
label=bwlabel(cropbw1,4);
stat=regionprops(label,'Centroid','Area','PixelIdxList','BoundingBox');
[maxValue,index] = max([stat.Area]);
[rw col]=size(stat);
for i=1:rw
    if (i~=index)
```

```matlab
        cropbw1(stat(i).PixelIdxList)=0; % Remove all small regions except large area index
    end
    if (i==index)
        alength=stat(i).BoundingBox(1,3)/Conversion;
        blength=stat(i).BoundingBox(1,4)/Conversion;
    end
end
figure,imshow(cropbw1)
imwrite(cropbw1,'6.jpg');
cropbw2 = bwconvhull(cropbw1,'objects');
label2=bwlabel(cropbw2,4);

stat2=regionprops(label2,'Area');
[maxValue_filled,index] = max([stat2.Area]);
figure,imshow(cropbw2)
imwrite(cropbw2,'7.jpg');

figure
imshow(crop);
img_mask = alphamask(cropbw2,[255 255 255],0.5);
saveas(img_mask,'8','jpg');
alength*1000
blength*1000
Area = maxValue_filled/(Conversion^2);
Volume = 0.0079/960/1000;
Thickness = Volume/Area
```